\documentclass{article}
\usepackage{ijcai26}

\usepackage{times}
\usepackage{soul}
\usepackage{url}
\usepackage[hidelinks]{hyperref}
\usepackage[utf8]{inputenc}
\usepackage[small]{caption}
\usepackage{graphicx}
\usepackage{amsmath}
\usepackage{amsthm}
\usepackage{booktabs}
\usepackage{algorithm}
\usepackage{algorithmic}
\usepackage[switch]{lineno}
\usepackage{multirow}
\usepackage{latexsym}

\usepackage{multirow}
\usepackage{tikz}
\usetikzlibrary{calc, fit, positioning, shadows}
\usepackage{dsfont}
\usepackage{pgf}
\usepackage{pgfplots}
\pgfplotsset{width=5cm,compat=1.9}
\pgfplotsset{every tick label/.append style={font=\tiny}}
\usepackage{amsthm}
\usepackage{amssymb}
\usepackage{cleveref}

\usepackage{pifont}
\usepackage{xspace}

\newcommand{\rex}{\textsc{r}\textnormal{e}\textsc{x}\xspace}

\newcommand{\xai}{XAI\xspace}

\newcommand{\commentout}[1]{}

\newcommand{\ie}{i.e.\xspace}

\theoremstyle{plain}

\theoremstyle{definition}
\newtheorem{definition}{Definition}

\theoremstyle{remark}

\theoremstyle{plain}

\newcommand{\freqrex}{\textsc{f}\textnormal{req}\textsc{r}\textnormal{e}\textsc{x}\xspace}
\newcommand{\fft}{$\mathcal{F}$\xspace}
\newcommand{\stft}{STFT\xspace}
\newcommand{\gtzan}{GTZAN\xspace}
\newcommand{\mca}{MCA\xspace}
\newcommand{\ravdess}{RAVDESS\xspace}

\newcommand{\pedro}{DistilHuBERT\_1\xspace}
\newcommand{\AST}{AST\xspace}
\newcommand{\sanchit}{DistilHuBERT\_2\xspace}
\newcommand{\atuaans}{Wav2Vec2\xspace}
\newcommand{\whisper}{Whisper\xspace}

\newcommand{\firdhokk}{Whisper\xspace}
\newcommand{\hubert}{DistilHuBERT\xspace}
\newcommand{\wav}{Wav2Vec2\xspace}

\title{I Guess That's Why They Call it the Blues:\\Causal Analysis for Audio Classifiers}

\author{
 David A. Kelly$^1$\footnote{Contact Author}\And
 Hana Chockler$^1$\\
 \affiliations
 $^1$King's College London\\
 \emails
 \{david.a.kelly, hana.chockler\}@kcl.ac.uk
}

\begin{document}

\maketitle

\maketitle

\begin{abstract}
It is well-known that audio classifiers often rely on non-musically relevant features and spurious correlations to classify audio.
Hence audio classifiers are easy to manipulate or confuse, resulting in wrong classifications. While inducing a misclassification is
not hard, until now the set of features that the classifiers rely on was not well understood.
In this paper we introduce a new method that uses causal reasoning to discover features of the frequency space that are sufficient and necessary for a given classification. 
We describe an
implementation of this algorithm in the tool \freqrex and provide experimental results on a number of standard benchmark datasets.
Our experiments show that causally sufficient and necessary subsets allow us to manipulate the outputs of the models in
a variety of ways by changing the input very slightly. Namely, a change to one out of $240$,$000$ frequencies results in
a change in classification $58\%$ of the time, and the change can be so small that it is practically inaudible. 
These results show that causal analysis is useful for understanding the reasoning process of audio classifiers and can be
used to successfully manipulate their outputs. 

\end{abstract}

\section{Introduction}

\begin{figure*}[t]
   % \centering
    %\hbox{\hspace{-0.5cm} \includegraphics[scale=0.8]{images/blues.pdf}}
    \includegraphics[scale=0.75]{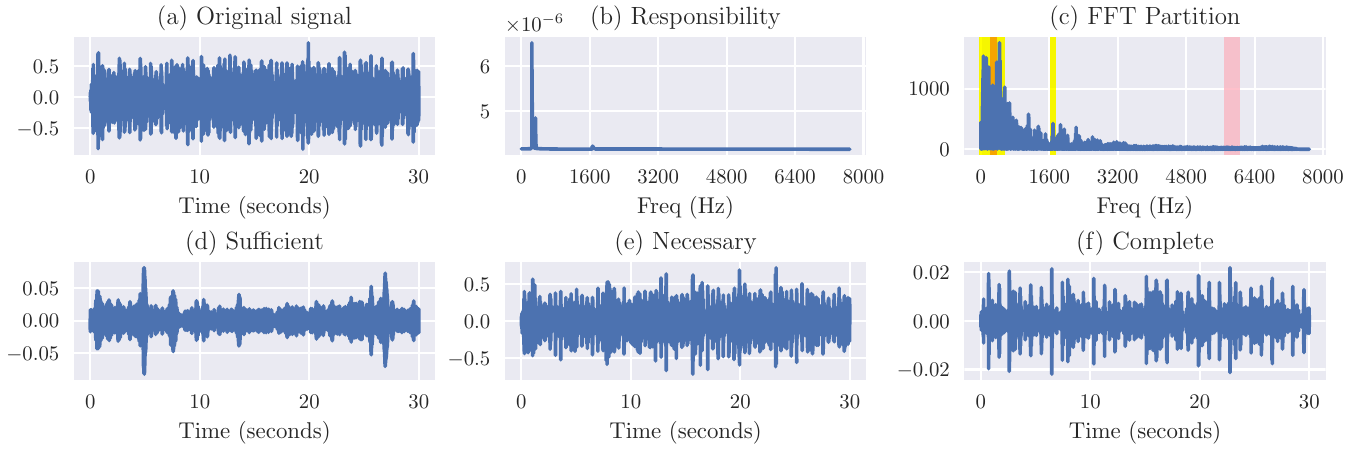}
    \caption{An example of \freqrex output. (a) shows the original signal, classified as `blues'. (b) is the \emph{responsibility} over \fft. (c) shows the sufficient frequencies (in orange), sufficient and necessary (combined orange and yellow) and finally, in red, the additional frequencies required to achieve the original model softmax score. (d) shows the reconstructed sufficient signal for the blues. This represents the bare minimum that the model will accept of this audio to be `blues'. (e) is the sufficient and necessary signal. Removing this from (a) gives the class `hiphop'. Finally (f) adjusts the model score of (e) \emph{down} towards the original score.}
    \label{fig:theblues}
\end{figure*}

Music content analysis (\mca) aims to build systems which work successfully with information in acoustic environments. One of the most common use cases is music genre recognition, or classification. This has many useful applications, most obviously in recommender systems.

Many recent advances in machine learning and AI, while not originating in \mca, have been successfully applied to this domain. The problem with such large, complex models is that they frequently use spurious -- non-musical -- features to make their classifications. \cite{sturm2014simple} likens these models to a $19^{th}$ century horse, Clever Hans, which appeared to be solving complex mathematical problems but was in fact just reading signals from the questioner. There are many ways to confuse the `horse' by blocking access to, or distorting, signal. But this does not tell us which part of a signal the horse was using, only that it can no longer use it correctly. 

We present \freqrex: a causality-based method and tool to decompose a signal into its sufficient, necessary and complete signal components. A \emph{sufficient}
signal is defined as a minimal subset of a signal required to have the same label as the original signal. A \emph{necessary} signal is both sufficient and necessary for the classification: by itself it has the desired class; removing it changes the class. We can capture the `essence of the blues' for a particular model by discovering the necessary signal for the `blues' classification. Finally, a \emph{complete} signal is a subset of the original signal which is sufficient, necessary and has the same confidence score as the original signal.

\freqrex's algorithms are based on actual causality~\cite{Hal19} to identify these different subsets of the signal. 
\freqrex is a frequency-based approach to signals, applying permutations to the Fourier Transform (\fft)~of the signal (\Cref{sec:freqrex}). 
As such, it needs no access to the model at all, beyond the ability to manipulate inputs and observe outputs --- it is entirely black-box. 

\freqrex allows us to isolate the most important frequencies for the model and manipulate them. Indeed, for all models, it is frequently possible to manipulate just one frequency out of the entire frequency space to change the classification. This manipulation may be just a simple alteration of the frequency's amplitude and phase. We also show that sufficiencies can be composed together and still be accepted by the model as having the correct classification, which no human would ever likely accept: as one would guess, they do not sound like music
at all, much less as music of a particular genre (\Cref{sec:eval}).

To the best of our knowledge, \freqrex is the first use of actual causality in audio, and the first tool to separate a signal into sufficient, necessary and complete components. We demonstrate \freqrex on $8$ different models across $2$ popular datasets, one for music genre classification and one for sung emotion classification. 
Knowing the cause of an outcome should permit us to manipulate that outcome:
we demonstrate the accuracy and, more importantly, utility of \freqrex's signal partitioning by showing that the classification can often be changed by manipulating only a subset of the sufficient frequencies.  
More precisely, we show that (1) combining multiple sufficient subsets results in audio that still has the same classification as the original subsets, even though it does not bear any musical resemblance to it; (2) using the subsets above, we can manipulate just one frequency out of $\approx240,000$ in a given input to change the classification with a $58\%$ probability of success, and manipulating
$5$ frequencies successfully changes the classification in $78\%$ of the inputs; and (3) using the Short-Time Fourier Transform we can introduce even smaller manipulations to the input (to the extent of not being audible at all) that still successfully change the classification in $62\%$ of the inputs.

All code, results, selected audio, and reproducibility material can be found in the supplementary material.

\section{Background}\label{sec:background}
\freqrex uses actual causality~\cite{Hal19} to analyse audio signal. In digital signal processing (DSP),
a system is considered causal if it is \emph{non-anticipative}; its present state is not influenced by its future state~\cite{oppenheim2013}. To clarify, we are not using the word `causality' to mean non-anticipative; rather, we apply the
existing theory of actual causality to the audio setting.

% While the full theory of actual causality is complex, we do not need most of it for our purposes. Essentially, the definition of actual cause is a generalization of `but-for' causality: if not for this, the outcome would have been different. A generalized notion
% of an \emph{actual cause} includes, in addition to the but-for causes, also inputs that contribute to the outcome, while not causing
% a change in it by themselves.

Actual causality is used in the eXplainable AI (\xai) tool \rex\footnote{\url{https://github.com/ReX-XAI/ReX}}. \rex~\cite{chockler2024causal} (an extension of~\textsc{DeepCover}~\cite{chockler2021explanations}) discovers subsets of pixels which make different contributions to a model's classification. For \rex, a `sufficient explanation' of an image is a minimal set of pixels which, by itself, is enough to have the same top-$1$ classification as the original. It is clear that an image may have more than one sufficient explanation~\cite{CKK25}. 
%\rex also finds sufficient and necessary pixels: those pixels which, by themselves, have the desired class and which, if removed, change the class. Finally, \rex also finds `complete' pixels; those pixels which change the softmax score of the sufficient and necessary pixels to match the original confidence score of the model~\cite{kelly2025causal}. For the remainder of the paper, we will use necessary to mean both sufficient and necessary and will clarify our usage if it is ambiguous.

\rex is an occlusion-based method which makes interventions over pixels and ranks them by their approximate \emph{causal responsibility}~\cite{CH04} to produce a \emph{responsibility map}. It uses this ranking to discover the aforementioned pixels sets. We avoid using the word `explanation' in the remainder of the paper, as it carries with it notions of (human) interpretability which are not relevant. 
Instead, we show that knowing causes allows us to manipulate the model. As ~\cite{bhusalface} state, `models are not constrained to use human-understandable cues; they only use features that minimize loss'. With that said, all of our output can be listened to, and therefore may form a basis for model explainability. 

\freqrex works over the frequency domain of the signal. We assume the signal is periodic, as required for the Fourier transform. We assume that both the Fourier and Short-Time Fourier transforms are well defined over the data. 

\paragraph{Fourier Transform \fft.} For a given signal $f(x)$, 
\[ \mathcal{F}(\xi) = \int_{-\infty}^{\infty} f(x) e^{- 2 \pi \xi x} dx \, \forall \xi \in \mathbb{R} \] 
Multiplication (denoted $\cdot$) and convolution (denoted $*$)
transform into each other, i.e. $\mathcal{F}[ s * r] = \mathcal{F}[s] \cdot \mathcal{F}[r]$. The Fourier transform $\mathcal{F}$ is invertible under certain conditions. We denote the inverse of \fft by $\mathcal{F}^{-1}$. Finally, we denote the Short-Time Fourier transform (STFT) by $\mathcal{F}_T$ and its inverse $\mathcal{F}_T^{-1}$. See~\cite{oppenheim2013} for a detailed presentation of the Fourier transform.

\section{Causality in Audio}\label{sec:audio_cause}

We adapt the definitions given for images~\cite{chockler2024causal,kelly2025causal} to apply to the frequency space of the signal. We refer the reader to those papers for the details on the structure of the underlying causal model and justifications for the algorithm.

Given a classifier model $\mathcal{N}$, a signal $x$ and its Fourier transform $\mathcal{F}(x)$, we define a sufficient subset as follows.

\begin{definition}[Sufficient Subset]\label{def:sufficient}
    A sufficient subset is a subset $S \subseteq \mathcal{F}(x)$ such that $\mathcal{N}(\mathcal{F}^{-1}(S)) = \mathcal{N}(x)$ and that $S$ is minimal, \ie there is no strict subset $S' \subset S$ such that $\mathcal{N}(\mathcal{F}^{-1}(S')) = \mathcal{N}(x)$. 
\end{definition}
\noindent
 Here $\mathcal{N}(x)$ refers to the top-$1$ class prediction, not the entire output tensor, or score. We use $\mathcal{N}_R(x)$ to denote both the top-$1$ class and its associated real value score.

 The following definition quantifies the contribution of a feature to the classification and is loosely inspired by the definition of responsibility in actual causality~\cite{CH04}.
 \begin{definition}[Sufficient responsibility]\label{def:resp}
 A frequency $s$ has a \emph{sufficient responsibility} (`responsibility' in short)
 $1/|S|$ for the classification $\mathcal{N}(x)$, where $S$ is a smallest sufficient subset $S \subseteq \mathcal{F}(x)$ 
 as defined in \Cref{def:sufficient}, such that $s \in S$. If there is no such $S$, the responsibility of $s$ is defined as $0$.
\end{definition}

When the sufficient subset $S$ is translated into the time domain, we refer to it as the \emph{sufficient signal}.
A \emph{sufficient and necessary} subset is a smallest set of frequencies such that they are, by themselves, sufficient for the original classification, and removing them from the original signal while keeping everything else the same, changes the classification. We refer to this subset as the necessary signal 
when in the time domain.

\begin{definition}[Necessary Subset]\label{def:necessary}
    A necessary subset is a subset $N \subseteq \mathcal{F}(x)$ such that $\mathcal{N}(\mathcal{F}^{-1}(N)) = \mathcal{N}(x)$ and 
    $\mathcal{N}(\mathcal{F}^{-1}(\mathcal{F}(x) \setminus N)) \neq \mathcal{N}(x)$ and that $N$ is minimal, \ie there is no strict subset of $N$ which satisfies this requirement.
\end{definition}

\begin{definition}[Complete Subset]\label{def:complete}
    A complete subset is a subset $C \subseteq \mathcal{F}(x)$ such that C is sufficient, necessary and $\mathcal{N}_R(\mathcal{F}^{-1}(C)) = \mathcal{N}_R(x)$.
\end{definition}

A complete signal is a subset of the original signal which captures its main characteristics as determined by the model. 
Henceforth, unless it leads to ambiguity, we refer to just that subset $C \setminus N$ as the complete signal, as it `completes' the necessary signal $N$. 

The definitions in~\cite{chockler2024causal} assume \emph{causal independence} between the input variables. Assuming causal independence means 
we are free to intervene on one frequency without having to propagate the intervention to other frequencies. 
It is clear that in the real world, frequencies can be causally related. In the simplest case, the overtone series of a musical instrument is caused by the resonant body of the instrument reacting to a fundamental frequency.
However, digital signals are \emph{data} and can therefore be manipulated in ways that the real world, analogue signal, cannot. The assumption of causal independence --- but not correlation independence --- is reasonable on data and not unique to our approach. 
%We adapt \cite{bottleneck}, who originally discussed concepts: `we emphasize that we study interventions on the value of a \emph{variable} within the model, not on that \emph{variable} in reality.'

\section{\freqrex}\label{sec:freqrex}
\Cref{algo:freqrex} shows the high-level algorithm for \freqrex. The method `calculate\_responsibility' computes an approximate responsibility of 
each frequency in the audio signal $x$ according to \Cref{def:resp}. It
is similar to that used by \rex in that it partitions the input space into $p$ parts and queries the model on different combinations of those parts. It further refines sets of frequencies with non-zero responsibility using the same partitioning strategy. 
Unlike \rex, which relies on a masking value to `hide' pixels, audio has a natural `neutral' value: removal of elements. 
%we are able to directly remove elements of the signal as we perform mutations over $\mathcal{F}(s)$. 
This reduces the risk of such signals being out-of-distribution (o.o.d.), unlike occluded images which are highly likely to be o.o.d. Moreover, we do not assume that the parts are made up of contiguous elements. 

\begin{algorithm}[t]
    \caption{\freqrex($\mathcal{N}, x)$}
    \label{algo:freqrex}
    \begin{flushleft}
        \textbf{INPUT:}\,\, 
        a model $\mathcal{N}$,  a audio signal $x$, partition size $p$, no. iterations $N$, a distance $\epsilon$
       \\ \textbf{OUTPUT:}\,\, a set of sufficient frequencies $S$,
       a set of sufficient and necessary frequencies $N$, a set of complete frequencies $C$\\
    \end{flushleft}     
        \begin{algorithmic}[1]
            \STATE yf $\leftarrow \mathcal{F}(x)$
            \STATE $\mathcal{R}_g \leftarrow \emptyset$
            \FOR{$i \in N$}
                \STATE $\mathcal{R}_i \leftarrow$ calculate\_responsibility(yf, $p$)
                \STATE $\mathcal{R}_g \leftarrow \mathcal{R}_g \cup \mathcal{R}_i$
                \IF{Earth\_Movers\_Distance($\mathcal{R}_i, \mathcal{R}_g) \leq \epsilon$}
                    \STATE \textbf{break}
                \ENDIF
            \ENDFOR
            \STATE $O \leftarrow$ order yf by $\mathcal{R}_g$
            \STATE bf, sufficient, necessary, complete $\leftarrow  \emptyset, \emptyset, \emptyset, \emptyset$ 
            \FOR{$o \in O$}
                \STATE bf $\leftarrow$ bf $\cup$ o
                \IF{$\mathcal{N}(\mathcal{F}^{-1}(\text{bf})) = \mathcal{N}(x)$}
                    \STATE sufficient $\leftarrow$ bf 
                \ELSIF{$\mathcal{N}(\mathcal{F}^{-1}(\text{bf})) = \mathcal{N}(x) \land \mathcal{N}(\mathcal{F}^{-1}(\text{yf} \setminus \text{bf})) \neq \mathcal{N}(s)$}
                    \STATE necessary $\leftarrow$ bf
                \ELSIF{$\mathcal{N}_R(\text{bf}) = \mathcal{N}_R(x)$}
                    \STATE complete $\leftarrow$ bf
                \ENDIF
            \ENDFOR
            \RETURN sufficient, necessary, complete
        \end{algorithmic}
\end{algorithm}
 
We compute the \emph{Earth Mover's Distance} (EMD)~\cite{emd} between $\mathcal{R}_g$ and $\mathcal{R}_i$ at the end of each iteration, before updating $\mathcal{R}_g$ with $\mathcal{R}_i$. The EMD captures the minimum cost of turning one distribution into the other. If it is low, the two distributions were similar in shape. Once the difference in EMD is sufficiently small and stable, as determined by the hyperparameter $\epsilon$, \freqrex breaks from the main loop. 

There is a subtle point related to the stopping condition, which is not addressed by \rex. \rex stops when it achieves a minimal sufficiency, \ie when the greedy algorithm first discovers the target classification. This is not a stable stopping condition, due to the non-monotonicity of many classifiers. It is often reasonable that a model alters its classification in the face of new evidence brought about by the addition of extra features. The termination condition of
\freqrex includes a stability criterion, called `chain length', which requires that the classification stays the same for a predefined number of iterations of the final for loop. For simplicity, we do not present this
in \Cref{algo:freqrex}. We also require that the classification meets a minimum percentage of the original model score.

\section{Manipulating and Attacking Sufficiencies}\label{sec:attacks}

\begin{algorithm}[t]
    \caption{Fourier Attack($\mathcal{N}, x, R, p, \Delta)$}
    \label{algo:fa}
    \begin{flushleft}
        \textbf{INPUT:}\,\, 
        a model $\mathcal{N}$,  an audio signal $x$,
        a responsibility map $R$, a partition $p$ of $\mathcal{F}(x)$, a vector of mutations $\Delta$
       \\ \textbf{OUTPUT:}\,\, an altered signal $x'$, number of frequencies changed $n$, mutation $\delta$
    \end{flushleft}     
        \begin{algorithmic}[1]
            \STATE $F \leftarrow $ sufficient freqs from $p$, ranked by $R$
            \STATE $n \leftarrow 0$ 
            \STATE $\delta_m \leftarrow \text{max}(\Delta)$ 
            \STATE yf $\leftarrow \mathcal{F}(x)$ 
            \WHILE{True} 
                \STATE yf $\leftarrow$ apply $\delta_m$ to freqs $0\dots n$ in F
                \IF{$\mathcal{N}(\mathcal{F}^{-1}(\text{yf})) \not= \mathcal{N}(x)$}
                    \STATE break
                \ELSE
                    \STATE $n \leftarrow n + 1$
                \ENDIF
            \ENDWHILE
            \FOR{$\delta \in \Delta$}
                \STATE yf $\leftarrow \mathcal{F}(x)$ 
                \STATE yf $\leftarrow$ apply $\delta$ to freqs $0\dots n$ in F
                \IF{$\mathcal{N}(\mathcal{F}^{-1}(\text{yf})) \not= \mathcal{N}(x)$}
                    \RETURN $\mathcal{F}^{-1}(\text{yf}), n, \delta$
                \ENDIF
            \ENDFOR
            \RETURN $\emptyset, \inf, \inf$
        \end{algorithmic}
\end{algorithm}

\Cref{algo:fa} details our procedure for finding frequency level perturbations of the classification. 
We start from a list of mutations to which we assume some ordering. For example, the mutations might be a series of amplitude scalars, or delay durations, where the ordering is the obvious one in both cases. We take the sufficient frequencies from $p$ and order them by their responsibility in $R$. 

We start by mutating the most responsible sufficient frequency with the maximum permitted mutation. If this does not succeed in changing the classification, we add in the next most responsible frequency at the same degree of mutation. We keep doing this until a classification change occurs, we exhaust our search budget, or we run out of sufficient frequencies. Once we have discovered the number of frequencies, $n$, which we need to alter, we attempt to minimise the mutation that caused that change. We do this by iterating through the various mutations from small to large while holding the number of frequencies altered the same.

\begin{algorithm}[t]
    \caption{Short Time Fourier Attack($\mathcal{N}, x, w, F, \delta)$}
    \label{algo:stfta}
    \begin{flushleft}
        \textbf{INPUT:}\,\, 
        a model $\mathcal{N}$,  an audio signal $x$,
        a frame size $w$, a set of sufficient frequencies $F$, a mutation $\delta$
        \\
        \textbf{OUTPUT:}\,\, an altered signal $x'$, number of slices changed $n$
    \end{flushleft}     
        \begin{algorithmic}[1]
            \STATE Zxx $\leftarrow \mathcal{F_T}(x, w)$
            %\STATE yf $\leftarrow \mathcal{F}(x)$ 
            \STATE $F' \leftarrow$ match frequency bins in $F$ to freqs in Zxx 
            \STATE R $\leftarrow$ ordering of Zxx frames by magnitude of $F'$
            \FORALL{r $\in$ R}
                \STATE Zxx $\leftarrow$ apply $\delta$ to freqs $F'$ in frames r
                \IF{$\mathcal{N}(\mathcal{F_T}^{-1}(\text{Zxx)}) \neq \mathcal{N}(x)$}
                    \RETURN i, $\mathcal{F_T}^{-1}$(Zxx)
                \ENDIF
            \ENDFOR
            \RETURN $\inf, \emptyset$
        \end{algorithmic}
\end{algorithm}

% \Cref{algo:fa} works by altering frequencies over the entire signal. This results in changes which are frequently clearly audible, as noise or distortion. We demonstrate another method which uses our causal analysis to alter classifications. This method results in changes which are, in general, smaller than those produced by~\Cref{algo:fa}.

\Cref{algo:stfta} details our method for applying changes over $\mathcal{F_T}(s)$.
It starts from the output of~\Cref{algo:fa}, in order not to duplicate initial discovery work. 
The frequency bins of the $\mathcal{F_T}(s)$ are not identical to those of \fft, as the frame size, $w$, cannot capture the full frequency space. We match the frequency bins in $\mathcal{F_T}(s)$ to their closest fits in $\mathcal{F}(s)$ and use these frequencies, $F'$ as the basis for alterations. 
This is, in general, an information losing process, as many frequency bins in $\mathcal{F}(s)$ may be mapped to $1$ frequency bin in $\mathcal{F_T}(s)$.

We obtain an ordering of the different frames of $\mathcal{F_T}(s)$ by amplitude of frequencies in $F'$. Finally, we iterate through the ordering, applying the successful mutation from~\Cref{algo:fa} to the frequencies in $F'$. We stop as soon as the classification changes or we exhaust our research budget. As an additional refinement, we also performed multiple loops of~\Cref{algo:stfta}, increasing the window size, $w$, each time \emph{iff} the previous window size had not brought about an in-budget class change.

\section{Evaluation}\label{sec:eval}
We evaluate \freqrex by using the sufficient, necessary and complete signals that it computes as a guidance for manipulating the input. We show that
using \freqrex output allows us to change the model classification by modifying only a very small part of the input, indicating, in particular, the
high accuracy of \freqrex outputs (in other words, we use the size of the modification as a proxy for the quality of \freqrex).

We consider two different model tasks: music genre classification and emotion classification. The benchmark set for the music genre classification is \gtzan~\cite{GTZAN}, which is the most popular dataset for audio classification. 
The dataset consists of $1000$ audio files in wav format. Each file is $30$ seconds long and is categorised into one of $10$ difference classes: `blues', `disco', `metal', `reggae', `rock', `classical', `jazz', `hiphop`, `country' or `pop'. This dataset has limitations~\cite{sturm2013gtzan}, but we are not concerned directly with the dataset quality, as we are not training or evaluating models. The second dataset we consider is a collection of sung audio files taken from \ravdess~\cite{ravdess}. This dataset consists of $1014$ wav files split over $24$ different actors with categories expressing emotions: `neutral', `calm', `happy', `sad', `angry', `fearful', `disgust', and `surprised'.

All models are publicly available at HuggingFace~\cite{wolf-etal-2020-transformers}\footnote{\url{https://huggingface.co/}}. We use $5$ different models on the \gtzan datasets, and $3$ on \ravdess. All models are fine-tuned versions of large baseline audio models. All experiments were run on A40 NVIDIA GPUs. We set the minimum chain length to be $5$ and the minimum acceptable confidence threshold for success to be $0.5$ of the original score.

In the analysis below, we focus on the music genre classifiers, introducing the emotion classifiers only when the results are noticeably different. 
The full results for the emotion classifiers are in the supplementary material.

\paragraph*{Patterns of responsibility}
\begin{figure*}
    \centering 
    \includegraphics[scale=0.75]{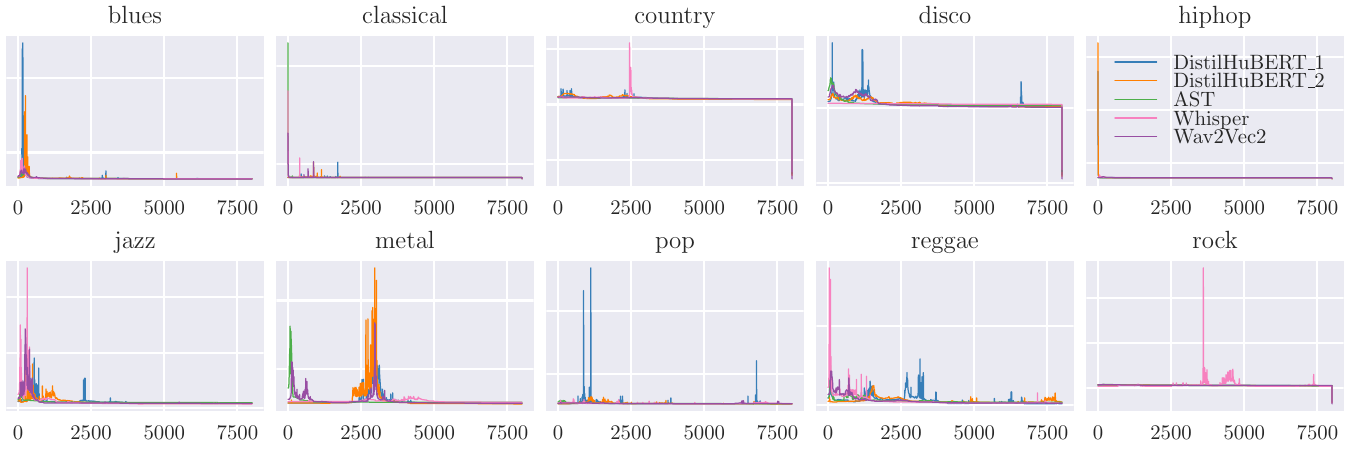}
    \caption{Average softmax of responsibility across $5$ models on the \gtzan dataset. While not identical, all models across most genres exhibit similar peak distributions, though the relative responsibility of the peaks varies per model. The $x$ axis is Hz.}
    \label{fig:average_resp}
\end{figure*}

\Cref{fig:average_resp} shows the average causal responsibility across the $10$ classes over $5$ models. The peak magnitudes (unit-less and therefore shown without specific $y$ values) are less important than the distribution of peaks. It clearly demonstrates similar behaviors of models for each genre,  suggesting that the models have all broadly learnt the same patterns in the data, regardless of their underlying architecture. Also, `country' and `disco' are obvious outliers in terms of their responsibility patterns, being similar to each other but different from all
other genres. It is not obvious why this is the case, as these two genres are quite dissimilar.

\paragraph*{Patterns of sufficiency and necessity}
\begin{table*}[t]
    \centering
    \begin{tabular}{l||r|r|r|r|r|r|r|r|r|r}
    \toprule
    \multicolumn{1}{c||}{Model} & \multicolumn{10}{c}{Genres} \\
    \midrule
       & blues & classical & country & disco  & hiphop  & jazz & metal  & pop & reggae & rock \\
      \midrule\midrule 
      \AST  &  3.68 & 0.37 & 20.42 & 7.93 &  4.67 & 9.64 &  1.81 & 4.81 & 1.83 & 21.90 \\
      \atuaans & 6.27 & 0.11 & 22.39 & 11.26 & 1.85 & 1.36 & 5.14 & 12.12 &  8.75 & 27.19\\
      \pedro & 1.22 & 0.44 & 20.29 & 17.78 & 0.30 & 6.80 & 1.87 & 1.98 & 10.86 & 40.39\\
      \sanchit & 2.03 & 2.22 & 15.36 & 21.45 & 0.11 & 9.45 & 0.68 & 6.25 & 9.78 & 27.25\\
      \whisper & 0.26 & 0.20 & 10.38 & 41.23 & 1.14 & 0.30 & 14.56 & 6.42 & 7.05 & 10.51\\
      \bottomrule
    \end{tabular}
    \caption{The average percentage (\%) of the frequencies in $\mathcal{F}(s)$ required for sufficiency across $5$ different models for the \gtzan dataset. In general, very little information is required.}%
    \label{tab:average_sufficiency}
\end{table*}

\Cref{tab:average_sufficiency} shows the average percentage (\%) of frequency space required by each model to classify each genre. It is immediately clear that, for example, `blues', `classical' and `hiphop' need significantly less information to be successfully classified, than `country' or `disco'. This is true across all models on the \gtzan dataset. 

\begin{figure*}[t]
    \centering
     \includegraphics[scale=0.75]{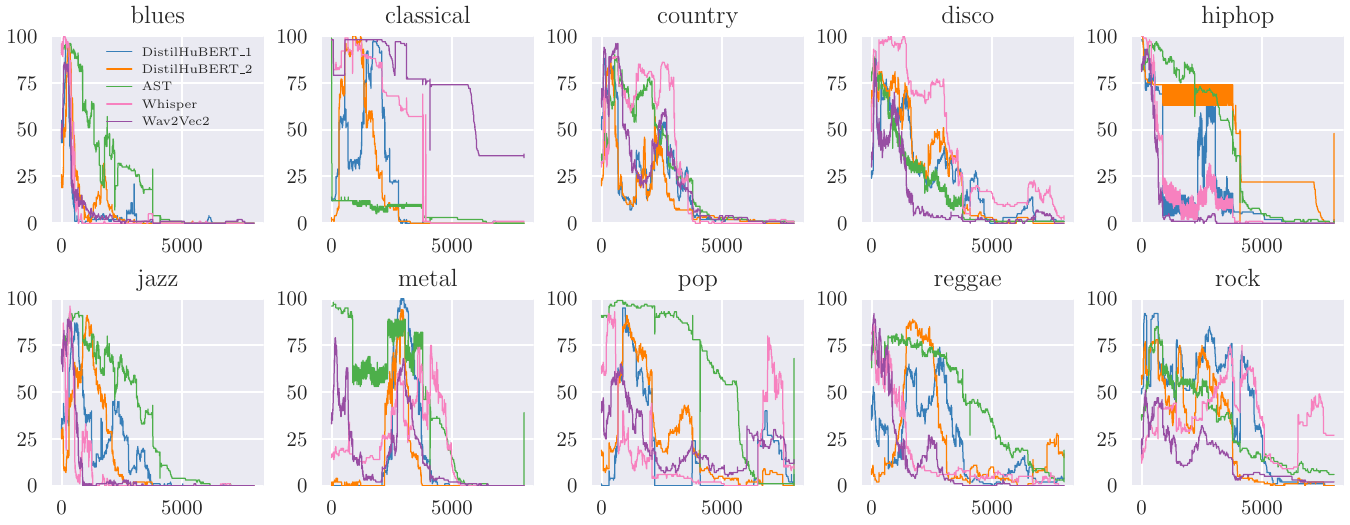}
     \caption{Frequency counts for necessary frequencies on \gtzan for correctly classified audio. While each model exhibits different patterns, there is a
     similarity between the curves of different models within the same genre. The $x$ axis is Hz.}
     \label{fig:average_sufficiency}
 \end{figure*}

\Cref{fig:average_sufficiency} shows how often different frequencies appeared in necessity signals over the $10$ classes of \gtzan. The general contours are broadly similar for each model over the different classes, with the exception of \AST, which is an obvious outlier on some classes. This is most noticeable on `pop'. Also curious is the small spike in the region of $8000$ Hz seen on `hiphop', `metal' and `pop' for \AST and \pedro. This frequency range is not usually considered to contain musical information, but the different models clearly consider it important for certain genres.

\paragraph*{Inversion Characteristics}
After subtracting the sufficient and necessary signal from the audio, there is usually enough signal remaining to query the classifier on the remnant. We refer to the classification of the `left over' signal as the inverse classification.
Reassuringly, both \pedro and \sanchit show similar inverse classification behavior, with `hiphop', `blues' and `reggae' dominating the inversions. They are not identical however, despite both fine-tuning the same underlying model. \pedro, for example, classifies the inverse signal as `hiphop' $409$ times, whereas \sanchit only $280$ times. The other architectures exhibit different characteristics, with `classical' dominating both \whisper and \atuaans, and \AST having a more uniform distribution over the inversions. 

\paragraph*{Completeness}
The complete signal adjusts the confidence score of the necessary signal so that it matches the score on the entire signal. Exact matching is very difficult, so we limit ourselves to $2$ decimal places. The complete signal can take the necessary score either up or down, and of course has a classification of its own.
Models \pedro and \sanchit have similar completeness characteristics, with a mean confidence score shift of $0.23$ ($\sigma = 0.17$), almost equally split between moving confidence up ($47\%$) and down. 

The \atuaans model has similar mean and $\sigma$, but the complete signal generally moves confidence score upwards, in $75\%$ of the cases. The final two models, \AST and and \whisper, have different patterns, with a lower mean score shift of $0.1$ for \AST and $0.18$ for \whisper. 

\subsection{Subset Manipulations}
We have shown that it is possible to capture and describe the different feature subsets of audio signal. We now show how to use this information, in a fully transparent manner, to bring about changes to model classifications.

\paragraph*{Sufficient Composition}\label{subsec:comp}
Given a collection of sufficient sets $s$ for class $c$ and model $\mathcal{N}$ it seems reasonable to see if $\mathcal{N}$ recognises the superposition of all sufficient signals with the same class $c$. This is a good test of how noise-free the sufficient signals are, and points at global characteristics that the models accepts as being class $c$. We present the results of this experiment in~\Cref{tab:compose}, with the results demonstrating that some genres are easier to produce than others. For example, `classical' has an extremely high composition ratio:  combining all $99$ correctly classified `classical' sufficient signals into one signal is still classified as `classical' by \AST, \atuaans and \sanchit. Conversely, no composition of `disco' was possible for \whisper. To a human ear, the composed signals do not sound like music of any particular genre (see the supplementary material). 

\begin{table*}[t]
    \centering
    \begin{tabular}{l||r|r|r|r|r|r|r|r|r|r}
    \toprule 
    \multicolumn{1}{c||}{Model} & \multicolumn{10}{c}{Genres} \\
    \midrule
       & blues & classical & country & disco  & hiphop  & jazz & metal  & pop & reggae & rock \\
      \midrule\midrule
      \AST  &  0.58 & 1.0 & 0.94 & 0.32 &  0.70 & 0.97 &  0.52 & 0.84 & 1.0 & 0.92 \\
      \atuaans & 0.57 & 1.0 & 0.90 & 0.80 & 0.49 & 0.95 & 0.98 & 0.75 &  0.61 & 0.89 \\
      \pedro & 0.99 & 0.98 & 0.70 & 0.97 & 0.98 & 0.88 & 1.0 & 1.0 & 0.79 & 0.94 \\
      \sanchit & 0.95 & 1.0 & 0.92 & 0.69 & 1.0 & 0.60 & 1.0 & 0.97 & 0.76 & 0.74 \\
      \whisper & 1.0 & 0.94 & 0.87 & 0.0 & 0.95 & 1.0 & 0.99 & 0.98 & 0.82 & 0.73 \\
      \bottomrule
    \end{tabular}
    \caption{Success rate of sufficiency superposition across genre and model.}\label{tab:compose}
\end{table*}

\paragraph*{Single Frequency Manipulation}\label{subsec:sfa}
Like image classifiers~\cite{supixel}, we find that audio classification models are also susceptible to failure modes based on a single feature.

\begin{table}[t]
    \centering
    \begin{tabular}{l||r|r|r}
         Model & Success & $1$ freq & $5$ freqs  \\ % & avg. shift \\
        \midrule\midrule
        \sanchit & $0.78$ & $0.48$ &  $0.57$ \\ % & $-698.45$ \\
        \atuaans & $0.55$ & $0.4$ & $0.69$  \\ % & $-467.89$ \\
        \pedro   & $0.79$ & $0.58$ & $0.72$ \\ % & $-668.62$ \\
        \whisper & $0.82$  & $0.25$  & $0.45$ \\ % & $-919.02$   \\
        \AST & $0.48$   & $0.18$  & $0.32$ \\ %  & $-703.97$   \\
        \midrule
        \firdhokk   & $0.99$ & $0.09$ & $0.20$ \\ % & $-622.22$ \\
        \hubert & $0.75$    & $0.35$  & $0.55$ \\ % & $-360.18$   \\
        \wav & $0.99$    & $0.79$  & $0.91$  \\ % & $-312.52$   \\
        \bottomrule
    \end{tabular}
    \caption{Success rate of frequency attacks on correctly classified audio. \wav is the easiest to attack with $1$ frequency, \AST the most resistant. Music genre is above the midline, emotion below.}\label{tab:freq_attack}
\end{table}

It is clear from~\Cref{tab:freq_attack} that both the music genre classifiers and emotion classifiers are highly susceptible to simple alterations of their sufficient frequencies. We set a resource budget of $1000$ alterable frequencies, with an amplitude shift as the permissible mutation. 
Considering that the total frequency space of most of the genre audio is 
$\approx 240$,$000$ for \gtzan, $1000$ is a very small number of permitted mutations -- only $0.4\%$. 
We find that the voice data is most easily manipulated by this approach, with \wav the most susceptible. It is possible to alter the classification $99\%$ of the time within the resource budget. $79\%$ of the successful attacks only required $1$ frequency to be altered by an amplitude shift. 

\begin{figure*}[t]
    \centering
    \includegraphics[scale=0.7]{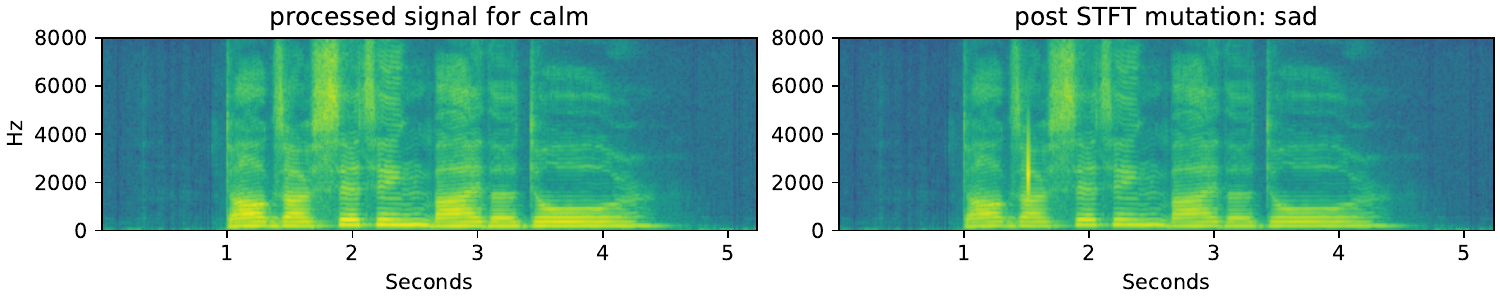}
    \caption{Two spectrograms, before and after \stft based alteration of the signal. There is a very small (imperceptible) change in amplitude between the two signal and a small noise spike around $1.6s$, yet this is enough to change the classification from `calm' to `sad'.}
    \label{fig:stft_happy}
\end{figure*}

\paragraph*{STFT Single Frequency Manipulation}\label{subsec:loc_sfa}
The results of~\Cref{algo:stfta} can be seen in~\Cref{tab:stft_success}. We consider only the audio which was successfully changed by~\Cref{algo:fa}. This is not a essential requirement of the algorithm but it makes the computation more efficient, as we already know what we need to do to succeed. 
\Cref{fig:stft_happy} shows an example on the \wav model where applying~\Cref{algo:stfta} resulted in a change of emotion from `calm' to `sad'. It is well-known that simple changes, such as amplitude or EQ settings, can affect classifiers~\cite{sturm2014simple}, but this method allows us both to compute a minimal change needed to the signal to affect the model and also understand why it works.

\begin{table}[t]
    \centering 
    \begin{tabular}{l|r||r|r|r}
        \toprule 
         & & \multicolumn{3}{c}{Frame Size} \\
        \midrule
         Model & Success & $256$ & $512$ & $1024$ \\
         \midrule\midrule 
       \AST  & $0.49$ &  $0.98$ & $0.2$ & $0.2$\\
       \sanchit  & $0.77$ & $0.97$ & $0.005$ & $0.017$ \\
       \atuaans & $0.56$ & $0.91$ & $0.03$ & $0.056$\\
       \pedro & $0.79$ & $0.97$ &  $0.003$ & $0.026$\\
       \whisper & $0.83$ & $1.0$ & $0.0$ & $0.0$ \\
       \midrule 
       \firdhokk & $0.99$ & $1.0$ & $0.0$ & $0.0$ \\ 
       \hubert & $0.76$ & $1.0$ & $0.0$ & $0.0$ \\ 
       \wav & $0.99$ & $1.0$ & $0.0$ & $0.0$ \\ 
       \bottomrule 
    \end{tabular}
    \caption{Total success and per frame size success for STFT manipulation. Music genre is above the midline, emotion data below.}\label{tab:stft_success}
\end{table}

%\subsection{Discussion}\label{subsec:discussion}
The question arises as to whether these STFT attacks are adversarial. In general, whether input is adversarial or not depends on the attacker model. The changes introduced to the signal in~\Cref{subsec:loc_sfa} are small alterations to floating point data. These changes frequently do not survive mapping back to the integer representation used by the wav audio format.  However, if we save the processed data directly as an array, then the transformation is robust. Given a HuggingFace `pipeline', it is easy to use these examples adversarially, as numpy arrays do not get preprocessed by the audio pipeline but are passed to the forward function without any translations or mapping. With an appropriate API, therefore, these results are adversarial.

\section{Related Work}
There is a number of \xai tools intended to explain model decisions in audio~\cite{siglime,mishra2017local,akman2024audio}. AudioLime~\cite{haunschmid2020audiolime} and MusicLime~\cite{sotirou2025} produce listenable output, as do we, but they rely on source separation. This lacks the fine control of our approach, which isolates individual frequencies, not individual voices or instruments. These methods will not work on \ravdess, as there is only one source in the signal. They do not partition the frequency space, nor demonstrate the utility of their method beyond human interpretability. 

There are many approaches to adversarial attacks on audio~\cite{carlini2018,du2020sirenattack}.~\cite{taori2019targeted} use a black-box method to achieve an attack success rate of $\approx 35\%$, lower than our success rate, but they make smaller changes and have a very different target. Our main goal is discoverability and utility of causes in audio, not adversarial attacks. Moreover, the way our method works is fully transparent.
One pixel attacks~\cite{supixel} showed that image classifiers can easily be perturbed by modulating a very small number of pixels. BlackCAtt~\cite{navaratnarajah2025out} uses causal reasoning, as do we, to change classifications, but they attack bounding boxes in visual object detectors. 

\cite{sturm2013classification} argues that the classification task is not valid when it comes to making a conclusion about whether a system is recognising genre, emotion, or some other semantic category. None of the systems we have investigated appear to use relevant musical knowledge. This problem exists also in image classification, where classifiers rely on very little information (e.g., $2\%$ of the input for transformers) in order to make their decisions~\cite{kelly2025big}. 
%Indeed, transformers in particular require only about $2\%$ of an image in order to provide a highly confident prediction.

\section{Conclusions and Future Work}
We have presented \freqrex, the first causal method for the analysis of sufficient, necessary and complete frequencies in audio classification. We have shown that it is possible to make composites of these frequency sets which maintain the original classification. We show that it is often enough to alter $1$ frequency to change classification outcome. In future work, we will use the sufficient, necessary, and complete sets of frequencies to improve adversarial attacks on the model and to fine-tune them in order to reduce their reliance on spurious features.

\section*{Acknowledgements}
Hana Chockler and David A. Kelly acknowledge support of the UKRI AI program and the Engineering and Physical Sciences Research Council 
for CHAI -- Causality in Healthcare AI Hub [grant number EP/Y028856/1].

\bibliographystyle{named}
\bibliography{all}

@article{sturm2014simple,
  title={A simple method to determine if a music information retrieval system is a “horse”},
  author={Sturm, Bob L},
  journal={IEEE Transactions on Multimedia},
  volume={16},
  number={6},
  pages={1636--1644},
  year={2014},
  publisher={IEEE}
}

@article{akman2024audio,
  title={Audio explainable artificial intelligence: A review},
  author={Akman, Alican and Schuller, Bj{\"o}rn W},
  journal={Intelligent Computing},
  volume={2},
  pages={0074},
  year={2024},
  publisher={AAAS}
}

@article{haunschmid2020audiolime,
  title={audiolime: Listenable explanations using source separation},
  author={Haunschmid, Verena and Manilow, Ethan and Widmer, Gerhard},
  journal={arXiv preprint arXiv:2008.00582},
  year={2020}
}

@inproceedings{wolf-etal-2020-transformers,
    title = "Transformers: State-of-the-Art Natural Language Processing",
    author = "Thomas Wolf and Lysandre Debut and Victor Sanh and Julien Chaumond and Clement Delangue and Anthony Moi and Pierric Cistac and Tim Rault and Rémi Louf and Morgan Funtowicz and Joe Davison and Sam Shleifer and Patrick von Platen and Clara Ma and Yacine Jernite and Julien Plu and Canwen Xu and Teven Le Scao and Sylvain Gugger and Mariama Drame and Quentin Lhoest and Alexander M. Rush",
    booktitle = "Proceedings of the 2020 Conference on Empirical Methods in Natural Language Processing: System Demonstrations",
    month = oct,
    year = "2020",
    address = "Online",
    publisher = "Association for Computational Linguistics",
    url = "https://www.aclweb.org/anthology/2020.emnlp-demos.6",
    pages = "38--45"
}

@ARTICLE{siglime,
  author={Abdullah, Talal Ali Ahmed and Zahid, Mohd Soperi Mohd and Turki, Ahmad F. and Ali, Waleed and Jiman, Ahmad A. and Abdulaal, Mohammed J. and Sobahi, Nebras M. and Attar, Eyad T.},
  journal={IEEE Access}, 
  title={Sig-Lime: A Signal-Based Enhancement of Lime Explanation Technique}, 
  year={2024},
  volume={12},
  number={},
  pages={52641-52658},
  doi={10.1109/ACCESS.2024.3384277}
}

@ARTICLE{supixel,
  author={Su, Jiawei and Vargas, Danilo Vasconcellos and Sakurai, Kouichi},
  journal={IEEE Transactions on Evolutionary Computation}, 
  title={One Pixel Attack for Fooling Deep Neural Networks}, 
  year={2019},
  volume={23},
  number={5},
  pages={828-841},
  doi={10.1109/TEVC.2019.2890858}}

@article{kelly2025causal,
  title={Causal identification of sufficient, contrastive and complete feature sets in image classification},
  author={Kelly, David A and Chockler, Hana},
  journal={arXiv preprint arXiv:2507.23497},
  year={2025}
}

@article{chockler2024causal,
  title={Causal explanations for image classifiers},
  author={Chockler, Hana and Kelly, David A and Kroening, Daniel and Sun, Youcheng},
  journal={arXiv preprint arXiv:2411.08875},
  year={2024}
}

@book{Hal19,
    author = {Joseph Y. Halpern},
    title  = {Actual Causality},
    year = 2019,
    publisher = {The MIT Press}
}

@inproceedings{CKK25,
   title={Multiple Different Explanations for Image Classifiers},
   author={Chockler, Hana and Kelly, David A. and Kroening, Daniel},
   booktitle = {{ECAI} European Conference on Artificial Intelligence},
   year = {2025}
}

@article{CH04,
  author    = {Hana Chockler and
               Joseph Y. Halpern},
  title     = {Responsibility and Blame: {A} Structural-Model Approach},
  journal   = {J. Artif. Intell. Res.},
  volume    = {22},
  pages     = {93--115},
  year      = {2004}
}

@misc{GTZAN,
    author = {Credit Fusion and Will Cukierski},
    title = {GTZAN},
    year = {2011},
    howpublished = {\url{https://www.kaggle.com/datasets/andradaolteanu/gtzan-dataset-music-genre-classification}},
    note = {Kaggle}
}

@article{sturm2013gtzan,
  title={The GTZAN dataset: Its contents, its faults, their effects on evaluation, and its future use},
  author={Sturm, Bob L},
  journal={arXiv preprint arXiv:1306.1461},
  year={2013}
}

@article{sturm2013classification,
  title={Classification accuracy is not enough: On the evaluation of music genre recognition systems},
  author={Sturm, Bob L},
  journal={Journal of Intelligent Information Systems},
  volume={41},
  number={3},
  pages={371--406},
  year={2013},
  publisher={Springer}
}

@inproceedings{kelly2025big,
  title={I am big, you are little; I am right, you are wrong},
  author={Kelly, David A and Chanchal, Akchunya and Blake, Nathan},
  booktitle={Proceedings of the IEEE/CVF International Conference on Computer Vision},
  pages={817--826},
  year={2025}
}

@inproceedings{mishra2017local,
  author       = {Saumitra Mishra and
                  Bob L. T. Sturm and
                  Simon Dixon},
  title        = {Local Interpretable Model-Agnostic Explanations for Music Content
                  Analysis},
  booktitle    = {Proceedings of the 18th International Society for Music Information
                  Retrieval Conference, {ISMIR} 2017, Suzhou, China, October 23-27,
                  2017},
  pages        = {537--543},
  year         = {2017},
}

@misc{sotirou2025,
      title={MusicLIME: Explainable Multimodal Music Understanding}, 
      author={Theodoros Sotirou and Vassilis Lyberatos and Orfeas Menis Mastromichalakis and Giorgos Stamou},
      year={2025},
      eprint={2409.10496},
      archivePrefix={arXiv},
      primaryClass={cs.SD},
      url={https://arxiv.org/abs/2409.10496}, 
}

@misc{ravdess,
  author       = {Livingstone, Steven R. and
                  Russo, Frank A.},
  title        = {The Ryerson Audio-Visual Database of Emotional
                   Speech and Song (RAVDESS)
                  },
  month        = apr,
  year         = 2018,
  publisher    = {Zenodo},
  version      = {1.0.0},
  doi          = {10.5281/zenodo.1188976},
  url          = {https://doi.org/10.5281/zenodo.1188976},
}

@inproceedings{bhusalface,
  title={FACE: Faithful Automatic Concept Extraction},
  author={Bhusal, Dipkamal and Clifford, Michael and Rampazzi, Sara and Rastogi, Nidhi},
  booktitle={The Thirty-ninth Annual Conference on Neural Information Processing Systems},
    year={2025}
}

@book{oppenheim2013,
  author   = {Oppenheim, Alan and Schafer, Ronald},
  title    = {{Discrete-Time Signal Processing }},
  pages    = {1056},
  publisher = {Pearson Deutschland},
  year     = {2013},
  isbn     = {9781292025728},
  doi      = {},
  url      = {https://elibrary.pearson.de/book/99.150005/9781292038155}
}

@inproceedings{chockler2021explanations,
  title={Explanations for occluded images},
  author={Chockler, Hana and Kroening, Daniel and Sun, Youcheng},
  booktitle={Proceedings of the IEEE/CVF International Conference on Computer Vision},
  pages={1234--1243},
  year={2021}
}

@INPROCEEDINGS{carlini2018,
  author={Carlini, Nicholas and Wagner, David},
  booktitle={2018 IEEE Security and Privacy Workshops (SPW)}, 
  title={Audio Adversarial Examples: Targeted Attacks on Speech-to-Text}, 
  year={2018},
  volume={},
  number={},
  pages={1-7},
}

@inproceedings{du2020sirenattack,
  title={Sirenattack: Generating adversarial audio for end-to-end acoustic systems},
  author={Du, Tianyu and Ji, Shouling and Li, Jinfeng and Gu, Qinchen and Wang, Ting and Beyah, Raheem},
  booktitle={Proceedings of the 15th ACM Asia conference on computer and communications security},
  pages={357--369},
  year={2020}
}

@inproceedings{taori2019targeted,
  title={Targeted adversarial examples for black box audio systems},
  author={Taori, Rohan and Kamsetty, Amog and Chu, Brenton and Vemuri, Nikita},
  booktitle={2019 IEEE security and privacy workshops (SPW)},
  pages={15--20},
  year={2019},
  organization={IEEE}
}

@article{navaratnarajah2025out,
  title={Out-of-the-box: Black-box Causal Attacks on Object Detectors},
  author={Navaratnarajah, Melane and Kelly, David A and Chockler, Hana},
  journal={arXiv preprint arXiv:2512.03730},
  year={2025}
}

@INPROCEEDINGS{emd,
  author={Rubner, Y. and Tomasi, C. and Guibas, L.J.},
  booktitle={Sixth International Conference on Computer Vision}, 
  title={A metric for distributions with applications to image databases}, 
  year={1998},
  volume={},
  number={},
  pages={59-66},
}

\end{document}